\documentstyle[prl,aps,psfig]{revtex}

\begin{document}

\title{Effects of Disorder on Ferromagnetism in Diluted Magnetic
Semiconductors}

\author{ Mona Berciu and R. N. Bhatt }

\address{Department of Electrical Engineering, Princeton University,
Princeton, New Jersey 08544}

\date{\today}

\twocolumn[\hsize\textwidth\columnwidth\hsize\csname@twocolumnfalse\endcsname

\maketitle

\begin{abstract}
We present results of a numerical mean field treatment of interacting
spins and carriers in doped diluted magnetic semiconductors, which
takes into account the positional disorder present in these alloy
systems. Within our mean-field approximation, disorder enhances the
ferromagnetic transition temperature for metallic densities not too
far from the metal-insulator transition. Concurrently, the
ferromagnetic phase is found to have very unusual temperature
dependence of the magnetization as well as specific heat as a result
of disorder. Unusual spin and charge transport is implied.
\end{abstract} 

 ]


Following the discovery of a ferromagnetic transition in
Ga$_{1-x}$Mn$_x$As at temperatures in excess of 100 K
\cite{Ohno1,Haya,Esch}, well above those found in counterparts based
on II-VI semiconductors \cite{Haury}, there has been a surge in
interest in the magnetic properties of diluted magnetic semiconductors
(DMS).  Theoretical models abound to explain the ferromagnetism
\cite{RKKY,MacD,Nagaev}.  While it is generally accepted that the
ferromagnetism is due to an effective interaction between the magnetic
ions (Mn) mediated by mobile carriers (holes, since Mn, a group II
element substitutes for Ga, a group III element), different models
differ in detail, {\em e.g.} whether the interaction is RKKY or not,
and also the approximations used to model the system.

In nonmagnetic doped semiconductors, such as phosphorus doped silicon
\cite{Paalanen1}, there has been no evidence for ferromagnetism due to
carriers. Indeed, carrier hopping at low doping concentrations in the
insulating phase is known to induce antiferromagnetic interactions
between localized states, leading to a valence-bond-glass like state
down to the lowest observable temperatures \cite{B-L}. In contrast,
ferromagnetic tendencies were detected in doped diluted II-VI magnetic
semiconductors already in the insulating regime at low temperatures
\cite{Liu}, and subsequently ferromagnetism was observed in both II-VI
and III-V semiconductors at metallic doping densities.

In insulating DMS, the presence of Mn has been shown\cite{W-B-D,A-B}
to overwhelm the antiferromagnetic interaction between charge
carriers,leading to an essentially ferromagnetic ground state.  Monte
Carlo simulations \cite{W-B} for II-VI DMS in the insulating phase
show that the ferromagnetic phase is very unusual, with a highly
inhomogeneous magnetic profile, leading to unconventional properties
such as M(T) curve that is not described by expansions around the
critical point (critical point theories) or zero temperature (spin
wave theories) over most of the ferromagnetic phase.  By contrast,
theoretical models for the metallic regime \cite{RKKY,MacD} have been
based on the homogeneous electron gas, with a few exceptions, such as
the possibility of phase separation \cite{Nagaev}.

It is well-known in conventional doped semiconductors that the carrier
wavefunctions are those derived from an impurity band, for densities
in the vicinity of the metal-insulator transition (MIT), up to factor
of 3-5 above the MIT density, $n_c$\cite{Fritzsche}. Density
Functional calculations\cite{RSS} for a lattice of hydrogen atoms show
this clearly. The variation of critical density with uniaxial stress
in doped Si and Ge\cite{Paalanen2} is in agreement with calculations
based on impurity band wavefunctions \cite{Bhatt1}. Local moments are
known to dominate the low temperature behavior in doped semiconductors
well into the metallic phase\cite{A-P,Paalanen1}, and the effect is
enhanced for compensated systems\cite{Hirsch}. Raman
measurements\cite{Jain} in doped Si and infrared spectra\cite{Drew} in
GaAs also show features in the metallic phase characteristic of the
impurity wavefunction.

Ferromagnetism is found in Ga$_{1-x}$Mn$_x$As not far from a MIT, with
insulating behavior seen both at low and high Mn concentration
\cite{Ohno3}. Further, the system is heavily compensated
\cite{Ohno3,Besch,Matsu} with a carrier density only around 10\% the
Mn density. (Mn is nominally an acceptor in GaAs and is expected to
donate one hole per Mn). The vicinity of the MIT, and the large
compensation, which implies large disorder, motivate studying a model
that takes into account disorder as well as the impurity potential at
the outset, to see what their effects are on the magnetic properties
of the system.

Given the added complications of disorder, we study a model based on
an impurity band of hydrogenic centers with spin-1/2 (instead of the
more complex s=3/2 wavefunctions appropriate for acceptors), coupled
to localized Mn d-electrons in their S=5/2 ground state. The s=3/2
case, while technically more complicated, should yield qualitatively
similar conclusions. The impurity band is described in terms of a
tight binding Hamiltonian of the ground state impurity wavefunctions
at the impurity sites, which are distributed randomly on the Ga
sites\cite{note0}. As in previous work \cite{W-B-D,W-B}, the carriers
are coupled to the Mn spins by an antiferromagnetic (AFM) Heisenberg
exchange interaction.  The Hamiltonian we study is thus:
\begin{eqnarray}
\label{eq1}
\nonumber {\cal H}&=&\sum_{i,j}^{}t_{ij}c^{\dagger}_{i\sigma}
c_{j\sigma} +\sum_{i,j}^{} J_{ij} \vec{S}(i)
\left(c^{\dagger}_{j\alpha} {1 \over 2} \vec{\sigma}_{\alpha\beta}
c_{j\beta}\right) \\ &-& g \mu_BH\sum_{i}^{}{\sigma \over 2}
c^{\dagger}_{i\sigma} c_{i\sigma} - \tilde{g} \mu_B H
\sum_{i}^{}S^z(i)
\end{eqnarray}
Here, $\vec{R}_i$ ($i=1,N_d$) denotes the random positions of the Mn
impurities, and $c^{\dagger}_{i\sigma}$ is the creation operator of a
hole with spin $\sigma$ in the bound state associated with the $i$th
Mn impurity.  The first term in Eq. (\ref{eq1}) describes the hopping
of holes between various site, with the hopping matrix $t_{ij}=t(\vert
\vec{R}_i-\vec{R}_j\vert)$ given by $t(r) = 2\left( 1 + r/ a_B\right)
\exp{(-r/ a_B)} \mbox{Ry} $ \cite{Bhatt1}, where the Ry is the binding
energy of the hole, $E_b$, and $a_B = \epsilon \hbar^2/m^* e^2$ is the
hydrogenic Bohr radius.  The second term is the AFM interaction
between Mn spins $\vec{S}(i)$ and hole spins. Since Mn spins are
strongly localized, the exchange integral is simply given by $ J_{ij}
= J \exp{ (-2 { \vert \vec{R}_i - \vec{R}_j \vert / a_B} ) }, $
reflecting the probability of finding the hole in the impurity state
around $j$ on the $i$th Mn spin.  The last line in Eq.  (\ref{eq1})
describes interactions with an external magnetic field H.

We study finite size lattices containing $L^3$ simple cubic unit cells
(lattice constant $a$) of the zinc-blende structure.  $N_d$ of the Ga
FCC sublattice are substituted at random by Mn, leading to a Mn
concentration $n_{Mn} = 4x/a^3$, where $x= N_d/4L^3$. The total number
of holes is $N_h = p N_d$, implying a hole concentration $n_h=p n_{Mn}
$.  In all simulations presented in this paper we choose $L$ such that
for the corresponding $x$ and $p$, we have $N_h >50$ and $N_d > 500$,
so as to minimize finite size effects. Thus, in the absence of
external magnetic fields, the problem can be scaled in terms of four
dimensionless parameters: $J/E_b$, $a_B/a$, $n_ha_B^3 $ and $x$.
                                      
In this paper, we use parameters believed to be appropriate for
Ga$_{1-x}$Mn$_x$As: lattice constant $ a=5.65\AA$, hole binding energy
$E_b=112.4$ meV=1Ry \cite{BG}, with a consequent Bohr radius (in our
model) of $a_B=7.8\AA$\cite{note1}, and an exchange integral $J=15$
meV\cite{note2}.  Typical values of the Mn and hole concentrations are
$x=0.01-0.05$ and $p=5-10\%$ \cite{Besch,Matsu}. A more comprehensive
study, including a number of effects left out of this model, is being
completed\cite{berciu}. With these parameters, the typical hopping
parameter is $t(4a_B)=20$meV, though it should be emphasized that
$t_{ij}$ are distributed over a wide range\cite{berciu}.

We treat the AFM interaction within the mean-field approximation
(MFA), which leads to the replacement $\vec{S}(i)\hat{\vec{s}}_j
\rightarrow \langle\vec{S}(i)\rangle \hat{\vec{s}}_j + \vec{S}(i)
\langle \hat{\vec{s}}_j\rangle - \langle\vec{S}(i)\rangle
\langle\hat{\vec{s}}_j\rangle$, where $\hat{\vec{s}}_j=
c^{\dagger}_{j\alpha}{ 1 \over 2}\vec{\sigma}_{\alpha\beta}
c_{j\beta}$. The charge carrier Hamiltonian now contains the hopping
term and an effective on-site interaction $\sum_{j}^{}
\epsilon_{\alpha\beta}(j) c^{\dagger}_{j\alpha} c_{j\beta} $, with
$\epsilon_{\alpha\beta}(j)={1 \over 2}\sum_{i}J_{ij}
\vec{\sigma}_{\alpha\beta}\langle \vec{S}(i) \rangle$, and can be
numerically diagonalized for any configuration
$\langle\vec{S}(i)\rangle$ of the Mn spins, allowing the calculation
of the charge carrier spin expectation values $\langle
\hat{\vec{s}}_j\rangle$.  In turn, these allow us to compute the new
expectation values for the Mn spins, $|\langle\vec{S}(i)\rangle|={\cal
B}_S(\beta H_i), \vec{S}(i) \parallel \vec{H}_i $, where
$\vec{H}_i=-\sum_{j} J_{ij}\langle \hat{\vec{s}}_j\rangle$ and ${\cal
B}_S(x)$ is the Brillouin function.

The process is repeated until self-consistency is reached at each site
\cite{berciu}. As usual in MFA, the symmetry to spin rotations is
spontaneously broken and the expectation values are non-zero for some
direction, which we choose

\begin{figure}
\centering
\parbox{0.5\textwidth}{\psfig{figure=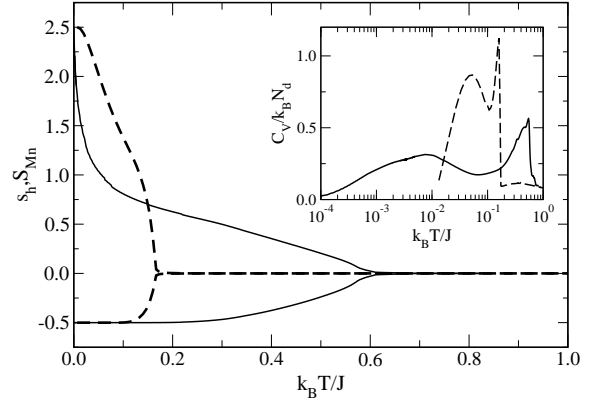,width=85mm,angle=270}}
\caption{\label{fig1} The average Mn spin $S_{Mn}(>0)$ and the average
spin per hole $s_{h}(< 0)$ for a typical random Mn distribution (full
lines) and a simple cubic ordered Mn distribution (dashed lines), both
for $x=0.00926$ and $p=10\%$ (see text). Inset shows the corresponding
specific heats per Mn spin. }
\end{figure}
\noindent 
\noindent as the z-axis (this is equivalent to having an
infinitesimally small magnetic field).  The average contributions to
the total magnetization of the Mn and hole spins are then proportional
to $S_{Mn}=1/N_d \sum_{i}^{}\langle{S}^z(i)\rangle $ and $s_{h} =
1/N_h \sum_{i}^{}\langle\hat{s}^z_i\rangle$. As might be expected,
below a temperature $T_c$, the system develops non zero expectation
values for the Mn and hole magnetizations, through hole-induced
alignment of the Mn spins. In Fig.\ref{fig1} we show the average Mn
and hole spins as a function of temperature, $T$, for a system with
$x=0.00926$ and $p=10\%$, for a typical random Mn distribution (full
lines). For comparison, we also show the corresponding results (dashed
lines) for a system with the same Mn concentration, but with Mn ions
arranged on a simple cubic lattice, with a (super)lattice constant
$a_L=a / (4x)^{1/3}=3a$.  Due to their AFM interaction, the two
expectation values have opposite signs, with the Mn spin saturating at
${5\over 2}$ and the hole spin saturating at $-{1\over 2}$ at low
temperatures. (The total magnetization $M(T)$ of the system has a
T-dependence similar to that of $S_{Mn}$, since Mn spins outnumber
holes ten to one, and also have a higher moment).

The first observation is that the magnetization of the disordered
system does not have the Brillouin-function shape typical for uniform
ferromagnets.  This is in part due to the small carrier density
relative to the Mn spin density; however, an even greater effect comes
from the wide distribution of exchange couplings and hopping
integrals, because of which many Mn spins do not order down to
extremely low T.  This is made clear in the inset for Fig.\ref{fig1},
which compares the specific heat for the two configurations on a
logarithmic scale for T: the disordered case shows a pronounced peak
at temperatures {\em well} below $T_c$, and significantly below its
lattice counterpart. Further, we find that the average Mn moment at $
T = 0.2T_c $ is well below the saturation value of 5/2, in accord with
experimental data \cite{Besch}, but in contrast with results obtained
from homogeneous electron gas models.

\begin{figure}
\centering
\parbox{0.5\textwidth}{\psfig{figure=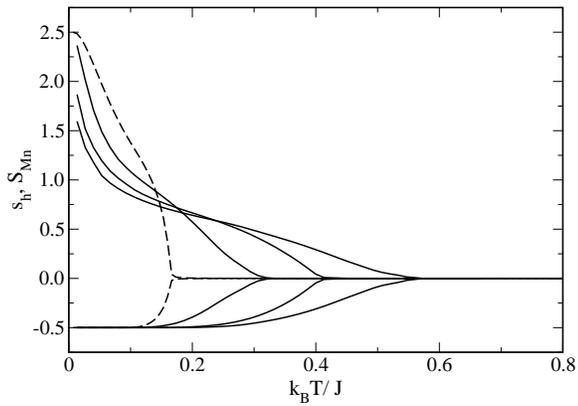,width=85mm,angle=270}}
\caption{\label{fig2} The average Mn spin $S_{Mn}$ and average spin
per hole $s_{h}$ for doping concentration $x=0.00926$ and $p=10\%$. In
increasing order of $T_c$, the curves correspond to an ordered, weakly
disordered, moderately disordered and completely random distributions
of Mn (see text).  }
\end{figure}

A second, surprising result is that randomness in the Mn positions
leads to a significant increase in $T_c$.  This is because, in the
disordered system, holes prefer regions of higher local concentration
of Mn, where they lower their total (magnetic and kinetic) energy, by
polarizing the Mn spins and hopping among several nearby Mn sites.  As
a result, these regions of higher Mn concentration become
spin-polarized at higher temperatures than in the uniform system, and
the resultant $T_c$ is increased. This is similar to a
percolation-like situation. We caution that this increase may be
significantly overestimated in a MFA such as ours, since spin
fluctuations between weakly coupled polarized clusters are not treated
accurately.

On the other hand, in a disordered system, the lower density Mn
regions have a lower than average probability to be visited by the
holes, and as a result the Mn spins in these regions only align
ferromagnetically at extremely low temperatures (see Fig.\ref{fig1});
such an effect is probably well captured by our scheme. We would like
to emphasize the fact that the holes at the Fermi energy at the
densities studied are either itinerant, or close to being so. An
analysis of hole wave-functions\cite{berciu} shows this
delocalization, along with the higher weight of holes in the regions
of high Mn concentration.  This delocalization is responsible for
alignment of the polarization at these high temperatures (relative to
the insulating system\cite{W-B}). Holes traveling between various
high-density regions force the alignment of Mn spins in each region to
be the same, in order to minimize their kinetic energy.

This picture can be checked by ``tuning'' the amount of disorder
(randomness) in the Mn positions. In Fig.\ref{fig2} we show the
average hole and Mn spins curves, for four types of Mn distributions,
with $x=0.00926$ and $p=10\%$.  In order of increasing $T_c$, they
are: (a) an ordered Mn cubic superlattice; (b) weak disorder,
corresponding to randomly displacing each Mn in (a) to one of the 12
nearest neighbor sites of the underlying FCC sublattice;
\begin{figure}
\centering
\parbox{0.5\textwidth}{\psfig{figure=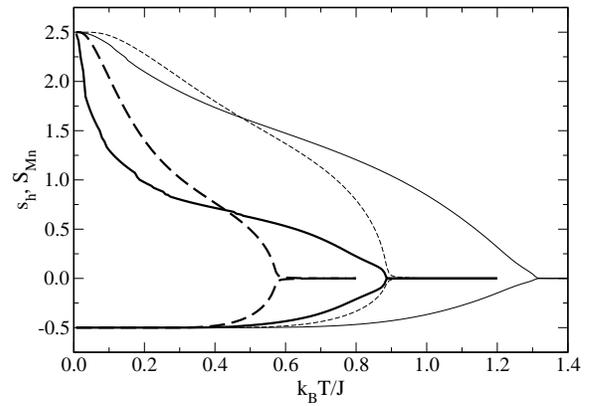,width=85mm,angle=270}}
\caption{\label{fig3} The average Mn spin $S_{Mn}$ and the average
spin per hole $s_{h}$ for doping concentration $x=0.05$ and $p=10\%$
(thick lines) and $30\%$ (thin lines) for typical random Mn
distributions (full lines) and simple cubic ordered Mn distributions
(dashed lines).  }
\end{figure}
\noindent (c) moderate disorder, corresponding to a random
distribution of Mn on the FCC sublattice, subject to a constraint that
all Mn-Mn distances are greater than $2a$; (d) completely random
distribution of Mn on the FCC sublattice. With increasing randomness,
$T_c$ increases, while saturation of $M(T)$ is simultaneously pushed
to lower $T$. The sensitivity to disorder suggests that carrier
density is not the only parameter characterizing the ferromagnetic
behavior in DMS; indeed, since original submission of this manuscript,
changes in M(T) curves with annealing time have been seen
experimentally\cite{samarth}.

We find a qualitatively similar picture holds for higher Mn
concentrations as well as higher hole densities, though the effects
are quantitatively less. (Our model, which does not include band
states, is likely to be less accurate at high densities).  In
Fig. \ref{fig3} we show the Mn and hole spins for both the simple
cubic superlattice and the random Mn distribution on the FCC
sublattice for $x=0.05$ for two different $p$.  While $T_c$ is again
larger in the random system in MFA, the percentage increase is smaller
than in the $x=0.00926$ case. Increasing the hole concentration from
$p=10\%$ to $p=30\%$ makes the curves more Brillouin-like.  This is
because the fluctuations in the local doping are smaller at higher Mn
concentrations, and increased hole doping further reduces the width of
the exchange distribution.

Our model, being based on the low doping limit, likely overestimates
the role of disorder; however, because ferromagnetism in DMS is seen
at low doping densities, not too far from the metal-insulator
transition, our work does strongly suggest that models based on the
homogeneous electron gas (whether mean-field\cite{MacD}, or with
perturbative RKKY exchange\cite{RKKY}) will not correctly capture the
nature of ferromagnetism. In particular, it casts doubt on their
quantitative fits to the observed $T_c$ in Ga$_{1-x}$Mn$_x$As. Our
calculation, while including the random positions of the Mn dopants,
leaves out disorder effects due to compensation, as well as
fluctuation effects left out in the MFA.  At higher Mn concentrations,
direct Mn-Mn interactions (which are needed to account for spin-glass
like behavior seen in many II-VI DMS above 20\% Mn) become important,
while for higher hole concentrations, one may have to include the band
states in addition to the impurity band, and possibly Coulomb
interactions between carriers as well. While these will have
quantitative effects on the results\cite{note3}, the unusual shape of
the magnetization curve and thermodynamic properties is likely to
remain at low doping, judging from the results of numerical Monte
Carlo simulations for the insulating phase of doped DMS\cite{W-B}.
Local experimental probes such as ESR and NMR would be especially
valuable in ascertaining any inhomogeneities in the magnetization and
carrier density profile, and help uncover the nature of ferromagnetism
in doped DMS at these low carrier densities.

In conclusion, the nature of ferromagnetism in doped DMS for low
doping, not too far above the metal-insulator transition density, is
strongly affected by disorder, which may, surprisingly, aid higher
$T_c$ in this regime. Further, by appropriate tuning of various
parameters, one may tailor the magnetic behavior M(H,T) in a manner
not possible in simple uniform magnets\cite{Kaneyoshi}.  This
versatility makes DMS ferromagnetism near the MIT a very interesting
problem from a theoretical point of view.  Adding to the richness are
possible effects of direct Mn-Mn interactions in concentrated systems
(which lead to spin-glass behavior in undoped II-VI
DMS\cite{Furdyna}), the existence of a ferromagnetic metal-insulator
transition (unlike conventional doped semiconductors and amorphous
alloys), and the likely unusual electron and spin transport
characteristics because of disorder.

This research was supported by NSF DMR-9809483. M.B.  acknowledges
support from NSERC, Canada.
R.N.B. acknowledges the hospitality of the Isaac Newton Institute for
Mathematical Sciences and of the Aspen Center for Physics while this
research was ongoing.


\begin{thebibliography}{99}
\bibitem{note0} Unlike Ga$_{1-x}$Mn$_x$As, where the hydrogenic
centers are located at the Mn spin sites, in II-VI semiconductors the
carrier dopants and Mn spins are independently distributed, leading to
substantial quantitative differences.
\bibitem{note2} The exchange integral is $J= 3 \epsilon$, where
$\epsilon= 5$ meV is the value obtained in Ref.  \cite{BG} for the AFM
interaction of a hole with the spin of its impurity Mn. We include the
factor of 3 as the simplest way to account for the fact that the heavy
holes have spin projections $j_z = \pm {3 \over 2}$, while in our
model they are modeled as $s_z= \pm {1 \over 2}$ objects.
\bibitem{note3} For this reason, the numbers obtained in this paper
are indicative, but not directly comparable with experiment.
\end{thebibliography}
\end{document}